\documentclass[runningheads]{llncs}

\usepackage[utf8]{inputenc} % allow utf-8 input
\usepackage[T1]{fontenc}    % use 8-bit T1 fonts
\usepackage{hyperref}       % hyperlinks
\usepackage{url}            % simple URL typesetting
\usepackage{booktabs}       % professional-quality tables
\usepackage{amsfonts}       % blackboard math symbols
\usepackage{nicefrac}       % compact symbols for 1/2, etc.
\usepackage{microtype}      % microtypography
\usepackage{graphicx,subfigure}
\usepackage{amsmath}
\usepackage[dvipsnames]{xcolor}
\usepackage{enumitem}
\usepackage{booktabs}       % professional-quality tables
\usepackage{floatrow}
\usepackage{adjustbox}
\newfloatcommand{capbtabbox}{table}[][\FBwidth]

\newcommand{\RV}[1]{\textcolor{black}{~#1}}

\usepackage{colortbl}
\newcolumntype{a}{>{\color{Sepia}}c}

\begin{document}
\newcommand{\system}{\textsc{ams}}
\newcommand{\mysection}[1]{\vspace{-2mm} \section{#1} \vspace{-.5mm}}
\newcommand{\mysectionn}[1]{\vspace{-2mm} \section*{#1} \vspace{-.5mm}}
\newcommand{\mysubsection}[1]{\vspace{-2mm} \subsection{#1} \vspace{-.5mm}}
\newcommand{\mysubsectionn}[1]{\vspace{-2mm} \subsection*{#1} \vspace{-.5mm}}
\newcommand{\cut}[1]{}

\title{Neural-based Modeling for Performance Tuning of Spark Data Analytics}%\thanks{Supported by organization x.}}
\author{Khaled Zaouk\inst{1} \and
Fei Song\inst{1} \and
Chenghao Lyu\inst{2} \and
Yanlei Diao \inst{1,2}
\authorrunning{K. Zaouk et al.}
%% First names are abbreviated in the running head.
%% If there are more than two authors, 'et al.' is used.
%
\institute{Ecole Polytechnique, France \and
University of Massachussets at Amherst, USA \\
\email{\{khaled.zaouk, fei.song, yanlei.diao\}@polytechnique.edu}
\\
\email{chenghao@cs.umass.edu}}}
\maketitle              % typeset the header of the contribution

%\begin{flushleft}

\vspace{-0.3in}
\begin{abstract}
	Cloud data analytics has become an integral part of enterprise business operations for  data-driven 
	insight discovery. Performance modeling of cloud data analytics is  crucial for performance tuning and 
	other critical operations in the cloud. Traditional modeling techniques fail to adapt to the high degree of 
	diversity in workloads and system behaviors in this domain.  In this paper, we bring  recent Deep 
	Learning techniques to bear on the process of automated performance 
	modeling of cloud data analytics, \RV{with a focus on Spark data analytics as representative  workloads}. 
	At the core of our work is the notion of learning workload embeddings 
	(with a set of desired properties) to represent fundamental computational characteristics of different jobs,  
	which enable  performance prediction when used together with job configurations that control 
	resource allocation and  other system knobs. 
	Our work provides  an in-depth study of different modeling choices that suit our requirements. 
	Results of  extensive experiments reveal %valuable insights including 
	the strengths and limitations 
	of different modeling methods, as well as  superior performance of our best performing method  
	over a state-of-the-art modeling tool for cloud analytics.
	
	%\keywords{First keyword  \and Second keyword \and Another keyword.}
\end{abstract}
\vspace{-0.3in}

\mysection{Introduction}

Big data analytics has become an integral part of enterprise businesses  for obtaining insights 
from  voluminous  data being generated every day. Big data analytics tasks %\RV{such as Spark-based tasks}, 
often run in the public cloud or  on  the enterprise's  private cloud. A  recent  survey reports that  currently, 65\% 
of North American enterprises rely on public cloud  platforms, and 66\% run  internal private 
clouds~\cite{forrester2020}.

Performance modeling of execution runs  (called jobs) of analytic tasks  on a cloud platform has become  
a critical  technical issue. From an analytical user's perspective, it is important to keep the {\em latency} 
of analytic jobs low in order to obtain timely insight from data, while at the same time, choose appropriate 
configurations (the number of cores, memory size, etc.) to reduce  {\em cloud  costs}, known to be a major 
part of operational expenses of companies today. From a cloud service provider's perspective, it needs to 
support  serverless computing (e.g., \cite{Aurora-serverless})  by estimating the latency of a  user 
job and deciding how  many resources to allocate  to offer the user a  cost-performance  sweet spot, 
as well as to use latency estimates to govern dispatching and admissions control~\cite{ChiMHT11}. 
Therefore, at the heart of massive-scale cloud analytics lies a critical technical issue: {\em estimate a performance  
objective (e.g., latency) of each analytic  job under a specific configuration (of resource allocation and other system 
knobs) on a cloud platform}, referred   to as the performance modeling problem in this paper. 
Then  the model can be used to tune  the  configuration in order to meet  cost-performance goals.

Performance modeling of cloud data analytics is a hard problem for several reasons: 
 First, there is a wide  spectrum of analytics tasks, such as SQL queries, user defined functions (UDFs), 
 machine learning (ML) tasks, that have different computational characteristics. Second,  analytics jobs are 
 run in a distributed  environment, involving highly complex, dynamic CPU, IO, memory, network  
 behaviors. Third, there are also many resource choices that affect performance (e.g., over 190 
  combinations of resource options covering the number of compute cores, memory, etc. 
  in  Amazon's EC2 offerings~\cite{EC2}). 
 For all of these reasons, existing modeling methods~\cite{LiDS15,RajanKCK16,Venkataraman:2016:EEP}  
 that employ manually-crafted models often fail to adapt to new analytic  workloads and resource options.

In this work, we bring large-scale machine learning to bear on the process of automated performance 
 modeling of cloud data analytics, \RV{with Spark analytics chosen as representative workloads that are 
 diverse in nature and widely deployed in the cloud}. 
 In this context, we explore the power of representation learning  of 
 Deep Neural Networks  (DNNs) to develop a {\em unified} solution to performance modeling, namely, 
 learning performance models of  analytic jobs solely from their runtime observations, irrespective of the 
 underlying task being SQL,  machine learning,  a mix of both, etc. At the core of our solution is the notion 
 of learning workload embeddings   for different jobs, 
 capturing their fundamental computational characteristics. Such embeddings,  
 when combined with a specific  configuration (of resources and other system knobs), 
 can be used to predict the performance of an analytic job on a cloud platform. 
  This modeling task, however,  
 is nontrivial   even for  DNNs due to complex real-world constraints in this problem domain, such as 
  the entanglement of different factors that affect performance and a limited number of observations 
  of an analytic task in training data (discussed in more detail in \S\ref{sec:system}.)
  Therefore, our work provides  an in-depth study of the different modeling choices 
  to suit the constraints of the problem domain, and reports on their effectiveness 
  for predicting latency of  both streaming and batch workloads  on top of Apache Spark~\cite{Spark} as 
  an example  distributed system.

More specifically, our contributions include the following:

\begin{itemize}
[nosep,leftmargin=1em,labelwidth=*,align=left]
\item 
We summarize complex constraints in real-world  cloud analytics applications and 
 outline our corresponding system design  (\S\ref{sec:system}).

\item 
We formulate the performance modeling problem, with an emphasis on learning workload embeddings 
to enable performance prediction. In addition, we leverage domain knowledge to propose three 
desired properties of such workload embeddings: \textit{reconstruction}, \textit{independence},
and \textit{invariance} to job configurations. Guided by these properties, 
we explore several families of modeling choices to learn both the workload embeddings and 
prediction models for complex analytic jobs. These choices include 
 ($i$) the embedding architecture; ($ii$) deep autoencoders augmented with customized 
 disentanglement;  ($iiv$) Siamese neural networks; ($v$) hybrid architectures (\S\ref{sec:modeling}).

\item
To enable large-scale evaluation, we collected runtime traces from an extension of a  
stream analytics benchmark \cite{LiDS15} as well as  the TPCx-BB benchmark~\cite{tpcxbb}, both of which 
 involve a mix of SQL queries and ML tasks. 
Results of extensive experiments show valuable insights, including the strengths and limitations 
of different modeling methods and superior performance of our best technique over a state of the 
art modeling method for data analytics~\cite{ottertune}. Most notably, 
Siamese networks, by offering the best approximation of the invariance property of embeddings,
enable the most accurate models to be learned.

\item End-to-end results with the best modeling technique, siamese networks, reveal reduction in runtime latency of $52.4\%$ on the streaming benchmark and $52.44\%$ on the TPCx-BB benchmark~\cite{tpcxbb}.

\end{itemize}

\mysection{Related Work}

We defer the discussion of relevant DNN models to Section~\ref{sec:modeling} 
where we present various modeling choices. Below we discuss 
a few broadly related areas of research.

\textbf{DBMS performance modeling and tuning.}
Prior work in the database community has addressed the problem of modeling and  tuning  database 
management systems (DBMS). OtterTune~\cite{ottertune,ottertune-demo} and CDBTune~\cite{Zhang:2019:EAC} 
are machine learning based solutions to performance tuning: they determine how to set the DBMS parameters 
by modeling a performance objective as a function 
 of the parameters and then iteratively exploring new configurations to update the model and move 
 the observed performance toward the optimum of the objective. As we  show in evaluation, 
 Ottertune \cite{ottertune,ottertune-demo},  by building a separate model for each job and mapping it to the 
 closest past job, offers inferior performance to our approach  grounded in  representation learning.  
 CDBTune~\cite{Zhang:2019:EAC} lacks the flexibility of returning 
 a performance model for any objective requested by the application.
Other performance modeling tools~\cite{LiDS15,RajanKCK16,Venkataraman:2016:EEP} use handcrafted models, 
and hence are hard to generalize. 
Recent work has used  neural networks  to predict latency of SQL query plans~\cite{Marcus:2019:PDN} or 
learn from existing query plans to generate better query plans~\cite{Marcus:2019:NLQ}.
These methods, however, are applicable  to SQL queries only, but not machine learning tasks or arbitrary UDFs.

\textbf{Cloud resource management.}
WiseDB~\cite{MarcusP16,MarcusSP17} proposes learning-based techniques for cloud 
resource management.  A decision tree is trained on a set of performance and cost related 
features collected from minimum cost schedules of sample workloads. 
Such minimum cost schedules are not available in our problem setting.
Paragon~\cite{paragon} and Quasar~\cite{quasar} cast the tuning problem into a collaborative filtering based 
recommender system. At the core of these systems are matrix factorization techniques that learn embedding 
vectors for both the workload as well as the configuration. As such, these systems do not allow to predict 
the performance over new configuration knobs, which stands in contrast
to our neural network recommender approach introduced later in Section~\ref{section: embedding-approach}

\textbf{Model search} tools such as  Hyperband~\cite{LiJDRT17} and 
Spearmint~\cite{SnoekLA12} aim to tune the hyperparameters of ML models. 
Spearmint, by using Bayesian Optimization as a core component, suffers from the cold-start problem: it requires several rounds of actively tuning the configuration for a submitted job before being able to recommend a good configuration. 
As a search-based tool, Hyperband tunes many configurations of hyperparameters by allocating increasingly more computation budget to more promising configurations. It, however, does not train predictive models that can guide efficient search for the best configuration.

\mysection{System Overview}
\label{sec:system}

In this section, we summarize requirements from real-world  use cases and outline our  system design. 

\textbf{Real-world requirements.}
We model an  analytic task as a  {\em dataflow program} (a directed graph of data collections flowing between  operations), which is used as the programming model in many systems like Spark~\cite{Spark},  Flink~\cite{Flink}, and Tensorflow \cite{tensorflow2015-whitepaper}. If the analytic task is a SQL query, we view the query plan returned by the query optimizer also as a dataflow program. A dataflow program is referred to as a {\em workload} in this paper.
For distributed execution, it needs to be transformed  to  a  cluster execution plan with resource allocation and other runtime knobs instantiated. When the  plan is executed, we call it a {\em job} and refer to all  runtime knobs collectively as the {\em job configuration}. In this work, we focus on four types of knobs (with examples given in Spark): 
(1)~\textit{resource allocation knobs}: e.g., the number of executors, number of cores per executor, memory per executor;
(2)~\textit{degree of parallelism}: e.g., the batch interval, block interval, parallelism;  
(3)~\textit{data shuffling}: e.g., the maximum size in flight, compression, bypass merge threshold;
(4)~\textit{SQL specific knobs}, as used in Spark SQL.

Practical use cases pointed to the following constraints and opportunities for performance modeling. 

1. {\em Generality for mixed workloads}. 
Analytics pipelines today  mix SQL queries for structured data analysis, SQL with user-defined functions (UDFs) for ETL tasks (data cleaning,  integration, etc.), and machine learning (ML)  tasks for deep analysis. In particular, UDFs and ML tasks are essentially black-box programs with computational characteristics unknown to the system. Given the diversity of analytic tasks, a general modeling approach  is needed to capture computational characteristics of these tasks. 
In our work, we use collected runtime metrics to do so, via {\em representation learning}. Note the distinction between the computational characteristics of a task (e.g.,  classifying the buying behaviors of users who received a coupon), and the job configuration (e.g., running the task with 10 cores, 2GB per core, using compression for data shuffling). The above two factors are orthogonal, but the collected runtime metrics reflect their {\em combined effect}.

2. {\em Limited observations per user task.}
A constraint in this problem domain is that for each user  task, there are only a small number of configurations that can be included in training data. This is because neither the service providers nor third-party entities (e.g., the optimizer) have the privilege to run user tasks outside their scheduled jobs, for which the user pays  the cloud cost. Hence, whenever a new job is submitted, we expect to have observed only one or a few (around 5) of its configurations.

3. {\em Offline sampling.} 
To overcome the issue of limited observations, an opportunity is that the modeling tool can use a separate benchmark, e.g., TPCx-BB used in our experiments~\cite{tpcxbb}, or a subset (e.g., 10\%) of client workloads  in the private cloud setting where the cloud is designed exclusively for a  client and hence the client is likely to offer some workloads to the system for sampling. For these workloads, called {\em offline workloads}, we seek to sample a large set of configurations using Bayesian optimization and heuristics based on Spark best practices.
Including such offline workloads in training helps develop an accurate model for {\em online workloads} (i.e.,  jobs that are triggered by the user application and incur cloud costs). It is because many real-world workloads are parameterized, i.e., generated from a set of common templates but with the parameters set to specific values by each application, and hence bear similarities across workloads.

\begin{figure*}
\begin{floatrow}
\hspace{-0.2in}
\ffigbox{
  \includegraphics[width=0.45\textwidth,height=3.8cm]{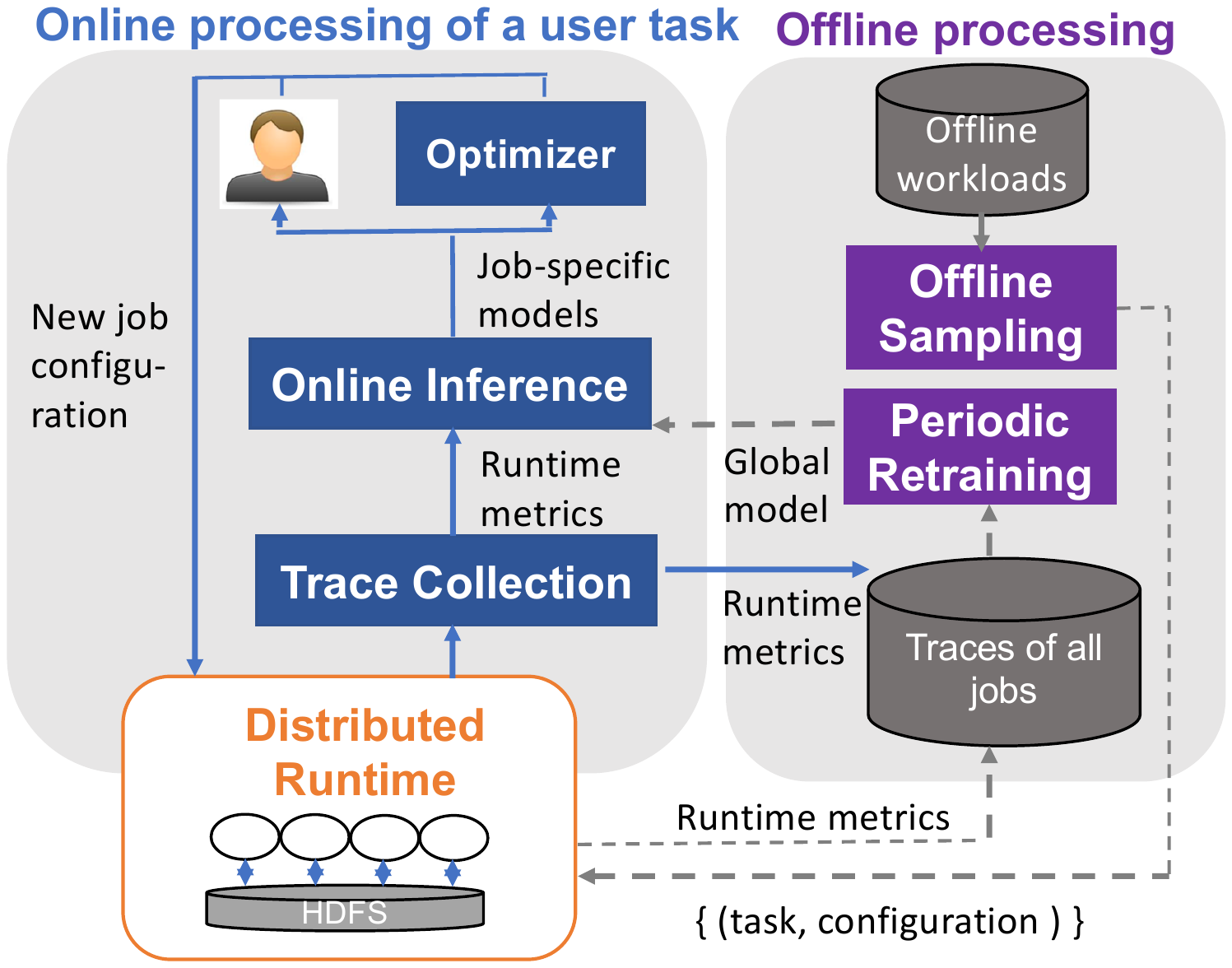}
  
}{
  \caption{Analytics Model Server (\system)}
  \label{fig:system}
}
\hspace{-0.2in}
\capbtabbox{
\footnotesize
\begin{adjustbox}{width=0.5\textwidth}
\begin{tabular}{cl} %{cl}
		\toprule 
		\textbf{Symbol} & \textbf{Description} \\ 
		\midrule
		$N$ & total number of training points\\
		%\hline
		$v_j^i$ & $i$th  configuration vector (of size $s$) set for job $j$ \\
		%\hline
		$\tilde{v}_j^i$ & approximation of  $v_j^i$ using an encoder \\
		%\hline
		$x_j^i$ & metrics vector (of size $p$) observed  from $v_j^i$ \\ 
		%\hline
		$\tilde{x}_j^i$ & approximation of  $x_j^i$ using auto-encoder\\
		%\hline
		$z^i_j$ & latent encoding vector (of size $k$) obtained   \\
		& for job $j$ using configuration $i$\\ 
		%\hline
		$z_j$ &  latent encoding vector (of size $k$) for workload $j$\\ 
		%\hline
		%	\hline
		$y_j^i$ & latency observed with the  configuration $v_j^i$    \\ 
		%\hline
		$f$ & regression fn outputing an approximation of  $y_j^i$\\ %maps knob configuration $v_j^i$ to
		%\hline
		$e / d$ & encoder / decoder function in  auto-encoder \\
		%\hline

		%	\hline
		\bottomrule
	\end{tabular}
\end{adjustbox}
}
{
  \caption{Notation}
  \label{tab:notation}
}
\end{floatrow}
\vspace{-0.1in}
\end{figure*}

\textbf{System design.}
The above requirements lead to our design of  a modeling system, called an Analytics Model Server (\system), as shown in Figure~\ref{fig:system}. 

The left panel  shows the {\bf online} path as a user job is submitted. 
The job is  run initially with a default or user-specified configuration. 
During job execution, \system\ collects a trace of metrics, collectively called an {\em observation},
including ($i$)~measures of performance objectives such as latency and cost;
($ii$)~engine-level metrics, e.g., Spark metrics such as time measurements of different steps, bytes read/written,  and fetch wait time; %bytes spilled to disk,
($iii$)~OS metrics such as CPU, IO, and network usage. 
As the metrics are collected, they are written to disk for persistent storage. 

The goal of modeling is to derive a job-specific prediction model, $f_j$, based on a global model trained using all past observations. 
If the workload is seen the first time, or its previous model was built from an outdated global model, the online inference module will use the current job observation to derive a new job-specific model from the most recent global model.  Then $f_j$ can be fed into an optimizer that automatically recommends a new configuration for the next execution in order to optimize the user objective (e.g., minimizing latency subject to a cost constraint). The above process repeats in future runs of the workload.

The right panel of Figure~\ref{fig:system} shows {\bf offline} processing with two modules. 
{\em Offline sampling}: \system\ uses heuristics from Spark best practices  or Bayesian optimization~\cite{SnoekLA12} to sample offline workloads, by selecting a wide range of  configurations and collecting their observations.
{\em Periodical retraining}: \system\  periodically retrains the global model by taking all past observations, including those from both online and offline workloads.

\mysection{Modeling Techniques}
\label{sec:modeling}
In this section, we  formulate the modeling problem and then present an in-depth study 
of  various modeling choices that suit our problem. 

\mysubsection{Formulating the Modeling Problem}
\label{sec:problem}

The main idea behind our modeling approach is that the performance of an analytic job 
(without loss of generality, we  use {\em latency} as an target objective in the following discussion)  is a function $f$ of
the computational characteristics of the workload and the job configuration (under fixed hardware infrastructure).
As stated in \S~\ref{sec:system}, the computational characteristics of a workload,
 for which we seek to learn a numerical representation called a  {\em workload embedding},
 are not known. 
At the core of our work is the notion 
of learning the workload embeddings automatically from runtime traces, and then combine the workload encoding 
and the job configuration to predict the latency of any arbitrary configuration of a given workload. If one bypasses 
the step of learning workload embeddings, the predictive power is limited, except when traces are generated from
configurations already within training data, and otherwise suffers from inferior performance, as our evaluation results shall show
in section \ref{sec:expt}.
 
Therefore, our modeling problem aims to learn (1)~the workload embedding $z_j$, for  job $j$, and 
(2)~a function $f^*$ that models  latency based on the embedding $z_j$, and a given $i$th configuration, 
denoted as $v^i_j$, of job $j$. The notation used in this paper is summarized in Table~\ref{tab:notation}.
In other words, we want to find both $f^*$ and $\{z_j\}$ such that:
\begin{equation} 
f^* = argmin_{f} \frac{1}{N} \sum_{i, j}  \left(f(z_j, v_j^i) - y_j^i \right)^2 
\end{equation}
where $N$ is the number of training points, 
$y_j^i$ is the observed latency of job $j$ under  configuration $v^i_j$, and 
 $\{z_j\}$ are latent embeddings that need be
learned from the observed runtime metrics $x_j^i$.

We propose a number of desired  properties of the embedding,  $z_j$, to guide the design of modeling: 

\begin{itemize}
[nosep,leftmargin=1em,labelwidth=*,align=left]	
	\item \textbf{Reconstruction}: For job $j$, the embedding $z_j$
	should allow the {\em reconstruction} of runtime metrics $x_j^i$ when coupled
	with the configurations $v_j^i$. 
	This is a typical property in learning representations from autoencoders.
	
	\item \textbf{Independence \& Invariance}: $z_j$ should be {\em independent} of
	 and {\em invariant} to the different configurations $\{v_j^i\}$ used for job $j$. These properties are   
	derived from  domain knowledge that the computational characteristics of a dataflow program 
	do not depend on, and further, do not vary with the resources used. We assume 
	that under fixed data characteristics it is possible to satisfy these properties, which in practice
	depends on the ability of the representation learning method to disentangle from runtime metrics 
	the characteristics of (parameterized) workloads  and the effect of the job configuration. 
	These properties, once achieved, will enable better accuracy in predicting latency 
	when an arbitrary (previously never seen) configuration is applied to a new workload. 

\end{itemize}

\begin{figure*}
	\vspace{-0.1in}
	
	\begin{tabular}{lcc}
		\subfigure[\small{Embedding architecture}]
		{\label{fig:embedding-architecture}\includegraphics[height=3.2cm,width=3.5cm]{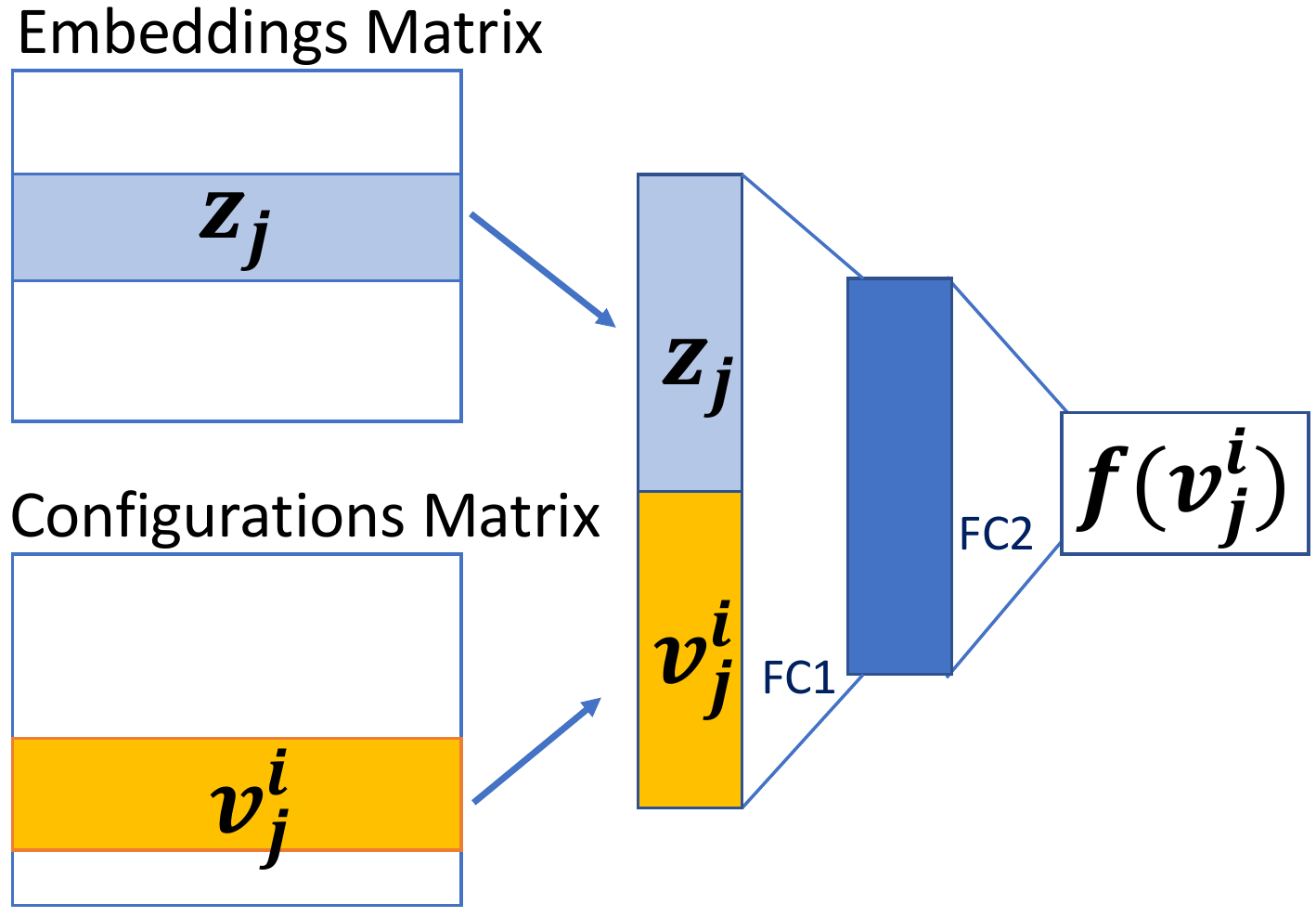}}

		&
		\subfigure[\small{Autoencoder architecture}]
		{\label{fig:autoencoder}
			\includegraphics[height=3.2cm,width=5.5cm]{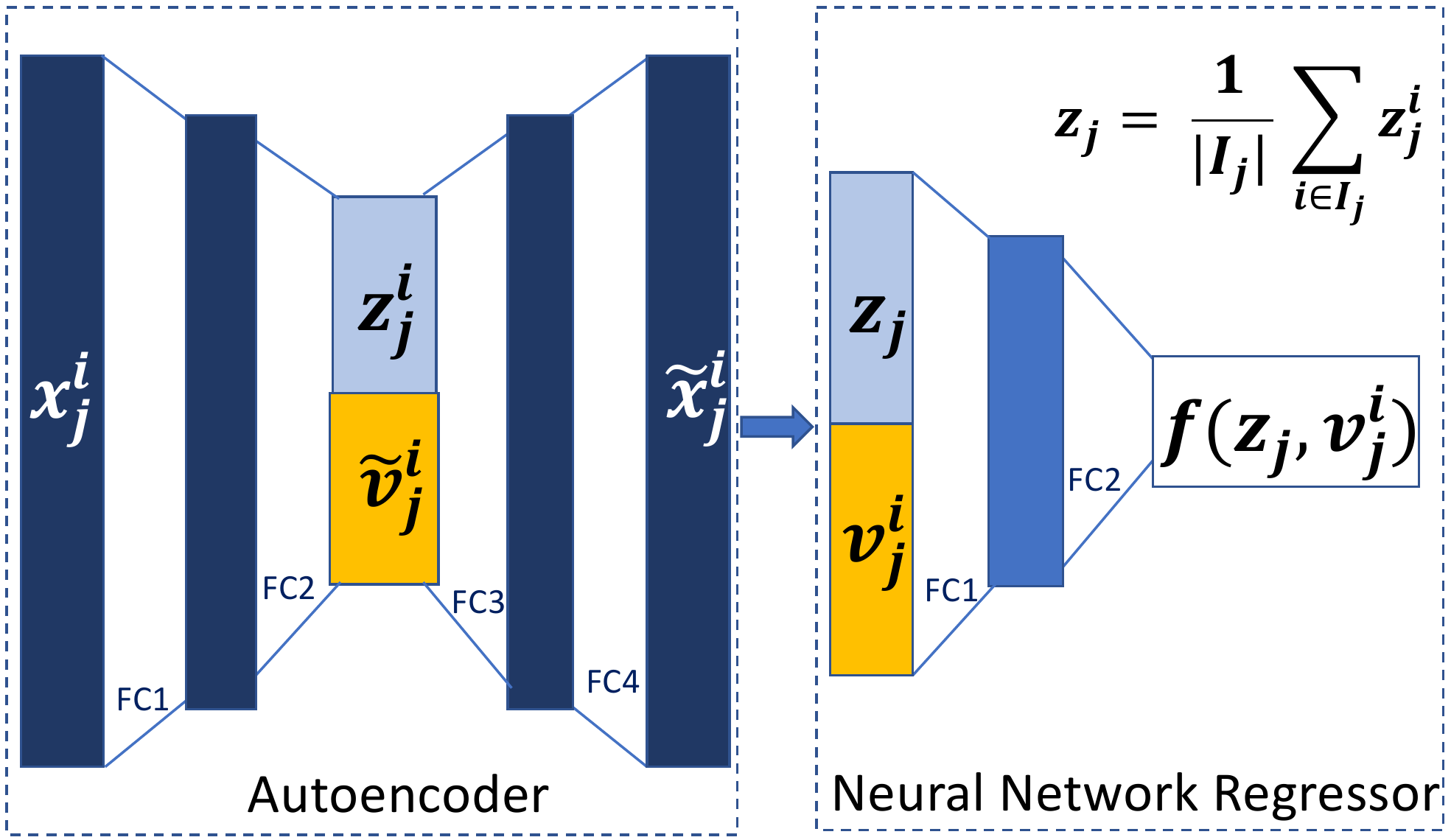}}
		
		&
		\subfigure[\small{Siamese neural network with triplets}]
		{\label{fig:siamese}\includegraphics[height=3.2cm,width=3.0cm]{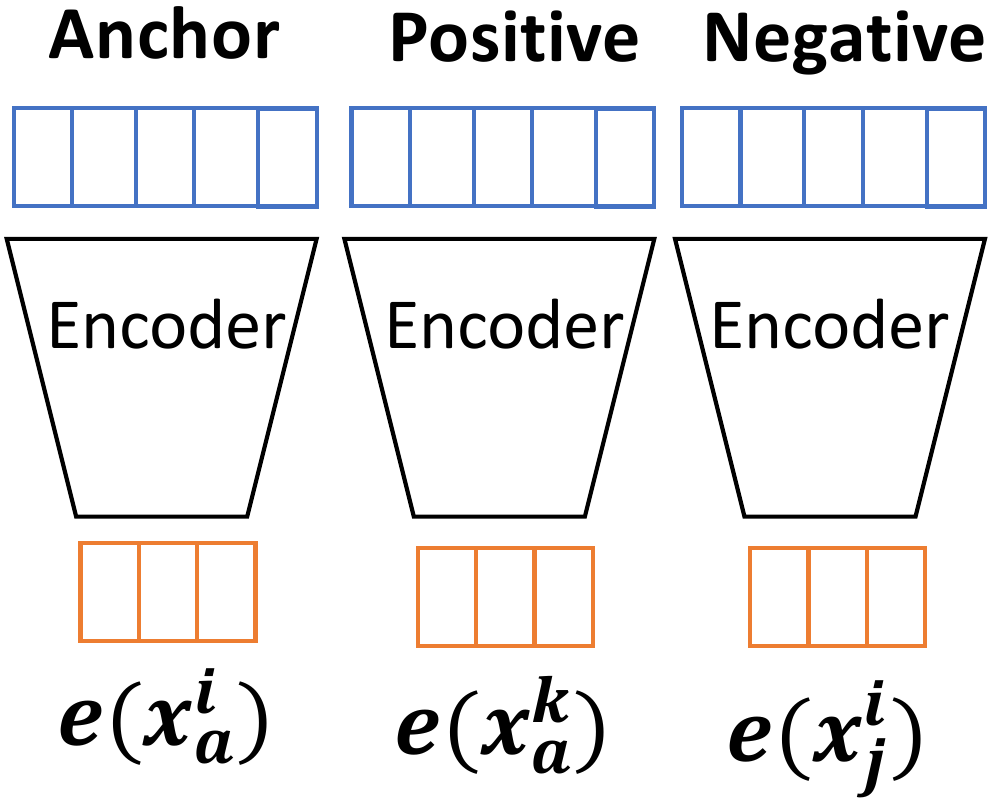}}

	\end{tabular}
	\caption{\small Three families of modeling choices.}
	\label{fig:setting_comparison}
	\vspace{-0.25in}
\end{figure*}

\mysubsection{Embedding Approach}
\label{section: embedding-approach}
The idea of the embedding architecture was inspired by deep recommender systems 
that embed user profiles in real-valued vectors while training the architecture to predict
user rankings of movies~\cite{He:2017:NCF:3038912.3052569}. By
analogy, we aim to embed workload characteristics in a real-valued
vector so that it can be used to predict latency of a particular workload.

The architecture, as shown in  Figure~\ref{fig:embedding-architecture}, couples 
representation learning and regression tasks within the same neural network. 
The architecture consists of three parts: 
(1)~An embedding layer with a weight matrix $Z$ denoting the latent space. Each
row $z_j$ of this matrix represents a particular workload $j$ as its embedding vector, 
randomly initialized first. 
(2)~A concatenation layer that for a particular job $j$, concatenates  the
embedding vector $z_j$  with an input ($i$th) configuration $v_j^i$ into,  $(z_j
|| v_j^i)$.
(3)~Several fully connected (FC) layers that take $(z_j || v_j^i)$ as the  input
and produce $f(v_j^i) \equiv f(v_j^i, z_j)$ as the final output.
The architecture is trained by minimizing the MSE between the predicted latency
$f(v_j^i)$ and the actual latency $y_j^i$, that is,    
$\frac{1}{N} \sum_{i,j} (f(v_j^i) - y_j^i)^2 $.

This architecture satisfies the independence and invariance properties since each
workload embedding is represented by a unique row vector within the embedding
matrix.
However, this approach requires incremental training every time a
new job is submitted: when the trace of a new job
becomes available, we add a random row to the embedding matrix, freeze the
weights of the neural network (except those at the embedding layer) and
run incremental training using the trace of the new workload and backpropagate to update
the embeddings.

\mysubsection{Encoder/Decoder based Approaches}

This family of approaches decouples workload extraction from the end 
regression task by using two neural networks, as shown in Figure~\ref{fig:autoencoder}.
A traditional autoencoder satisfies only the reconstruction property since it minimizes
the reconstruction loss while learning the encoding function: 
\begin{equation*} 
\mathcal{J} = \displaystyle\frac{1}{N} \sum_{i, j} ||\tilde{x}_j^i -
x_j^i||^2 
\end{equation*}
where $\tilde{x}_j^i$ denotes the approximation of the runtime metrics $x_j^i$  as 
output by the decoder. If we use $e$ to denote the encoding function, $d$ the 
decoding function, then $\tilde{x}_j^i = d(e(x_j^i))$.
 
Then the encoding in the bottleneck layer is fed to a neural network regressor to train a prediction model 
for latency. 
The regressor takes as input
the job configuration $v_j^i$ and $z_j$ (the centroid of $\{z_j^i\}_i$ for a
particular workload $j$) and tries to approximate at its output the runtime
latency $y_j^i$.

The loss function for the regression is simply the mean squared error:
\[ L = \displaystyle\frac{1}{N} \sum_{i, j} (f(v_j^i, z_j) - y_j^i)^2 \]
We can choose whether or not to fine tune the encoder layers while training the
downstream regression task.

{\bf Customized disentanglement.}
\label{custom-ae-paragraph}
Traditional autoencoders are not meant for explicitly disentangling data 
generation factors within the bottleneck layer. Thus, if we train a classical
autoencoder, the  bottleneck layer is unlikely to satisfy the independence
and invariance  properties stated above. Hence, we seek 
to guide the training of the autoencoder by adding domain knowledge and
explicitly breaking the bottleneck layer into two parts. The intuition here
is to force the encoding function to extract a variant part, $e_v(x_j^i)$, in a separated block 
of the bottleneck layer that tries to guess which configuration $v_j^i$ yielded the observation
trace given as input to the autoencoder. Then presumably, the other part of the bottleneck
layer, $e_{iv}(x_j^i)$, can become less variant for the traces coming from the same workload.
This architecture is depicted in Figure~\ref{fig:autoencoder}.
The loss function of the autoencoder with our customized disentanglement then balances
 the reconstruction term with a configuration approximation term:
\begin{equation}
\label{eq:custom-ae} 
\mathcal{J} = \displaystyle\frac{1}{N} \sum_{i, j} (||\tilde{x}_j^i -  x_j^i||^2 +
\gamma ||\tilde{v}_j^i -  v_j^i||^2) 
\end{equation}
where $\tilde{v}_j^i$ represents the encoder's approximation of the underlying
configuration when the input is $x_j^i$, and $\gamma$ is a regularization
coefficient.
In this setting, the encoder function $e$ is broken into two parts, $e(x_j^i)
= (e_v(x_j^i) || e_{iv}(x_j^i))$, where  $e_{v}(x_j^i) = \tilde{v}_j^i$ is an
approximation of the generating configuration, $e_{iv}(x_j^i) = z_j^i$
is the workload encoding, and  $d(e(x_j^i)) = \tilde{x}_j^i$.

{\bf Augmenting custom autoencoder with a contractive term.}
\label{custom-contractive-ae-paragraph}
We also augment our customized autoencoder by adding a Jacobian term to the loss
function, as introduced earlier in the literature of contractive autoencoders
\cite{DBLP:conf/icml/RifaiVMGB11}. Our intuition is to force the
designated invariant part of the  encoding, $e_{iv}(x_j^i)$, to become less variant to input
perturbations by adding the contraction term. 
\begin{equation}
\label{eq:contractive-ae}
\mathcal{J} = \frac{1}{N}
\displaystyle\sum_{i, j} \left( ||x_j^i - \tilde{x}_j^i||^2 + \gamma ||v_j^i - \tilde{v}_j^i||^2  + \lambda ||J_{e_{iv}}(x_j^i)||^2_F \right)
\end{equation}
where $J_{e_{iv}}$ is the Jacobian of the encoding output $z_j^i = e_{iv}(x_j^i)$ with respect to the input $x_j^i$.

\label{vae-paragraph}
\textbf{Variational autoencoders} \cite{Kingma2014AutoEncodingVB}
which belong to the family of generative autoencoders, are known for their ability to automatically
disentangle generating factors within the learned representations. 
The disentanglement effect comes from the independence assumption between different components of the posterior distribution of encodings, and is embodied by forcing the covariance matrix of this posterior distribution to be a diagonal matrix.
The loss function of $\beta$-variational autoencoders ($\beta$-VAE) \cite{Higgins2017betaVAELB}  balances between minimizing a reconstruction term and a KL divergence between the posterior distribution and the prior distribution. The reconstruction term indicates how much the distribution of encodings should trust the observed data, while the KL divergence term indicates how much this distribution of encodings should mimic the prior imposed on these encodings.
We compare the $\beta$-VAE to the previously introduced deterministic auto-encoders.

\mysubsection{Siamese Neural Networks}
\label{sec: siamese}
Interestingly, our problem is also related to the few-shot learning problem in object recognition since a new workload
is likely to have only a few observed configurations within training data. 
We thus propose in this section to use a Siamese network that has instead only an encoding part. This network aims to achieve the {\em similarity} property, as a relaxation of the invariance property. It encourages learning similar embeddings from different configurations corresponding to the same workload.
We first train this siamese network using a triplet loss~\cite{SchroffKP15}; we then introduce in the next section the
soft nearest neighbor loss~\cite{Frosst2019AnalyzingAI,pmlr-v2-salakhutdinov07a} as part of a hybrid architecture.

Training a siamese network with a \textbf{triplet loss} requires organizing the data (as shown in Figure~\ref{fig:siamese}) into triplets of:
%}
\begin{itemize}
[nosep,leftmargin=1em,labelwidth=*,align=left]
	\item Anchor point: $x_a^i$, which denotes the runtime metrics
	observed for an anchor job $a$ when the knob configuration is set to a
	particular value  $v^i$.
	
	\item Positive point: $x_a^k$, which denotes the runtime metrics
	observed for the same anchor job $a$ but with a different knob configuration
	$v^k$,  instead of $v^i$. 
	
	\item Negative point: $x_j^i$, which denotes the runtime metrics
	observed for a different job $j \neq a$ when the knob configuration is set
	to $v^i$, the same as the one used in the anchor point. 
\end{itemize}

At the input of the architecture, we provide 3 runtime metrics vectors: $x_a^i$,
$x_a^k$, $x_j^i$. The same fully connected layers are applied to
get the embeddings from the different observations, and we obtain their
respective embeddings: $z_a^i$, $z_a^k$, $z_j^i$. 
The  loss function on this instance of triplets is 
 $L_T(x_a^i, x_a^k, x_j^i)$ (defined below), and the final loss to be optimized is the sum over 
 all the instances of triplets:

\begin{equation*}
 	L_T(x_a^i, x_a^k, x_j^i)
 = \displaystyle
 max(0,||e(x_a^i) - e(x_a^k) ||^2 - ||e(x_a^i) - e(x_j^i)||^2 + \alpha)
 \end{equation*}
\begin{equation}
%\begin{adjustbox}
 \mathcal{J} = \displaystyle\sum_{a=1}^{n} \sum_{i=1}^{I_s} \sum_{j \neq a} 
 	L_T(x_a^i, x_a^k, x_j^i)
\end{equation}	
\noindent where $\alpha$ is a margin that is tuned alongside other hyperparameters.
The training of this loss function requires  $I_{s}$ shared
configurations across all  training workloads. However, an arbitrary configuration can be observed for the new workload at inference time.

\mysubsection{Hybrid Architectures} 
\label{section-hybrid-architectures}

In this section, we propose hybrid architectures that add decoders on top of Siamese neural networks.

\textbf{Hybrid1.}
We start by augmenting the previous architecture with a decoder in order to add to the triplet loss, additional terms related to our customized disentanglement and reconstruction. We thus minimize this loss function:
\begin{equation} 
\mathcal{J} = \displaystyle\sum_{a=1}^{n} \sum_{i=1}^{I_s} \sum_{j \neq a} L_T(x_a^i, x_a^k, x_j^i) + \gamma L_R(x_a^i, x_a^k, x_j^i) + \lambda L_C(v_a^i, v_a^k, v_j^i)
\end{equation}
with $L_T$ as provided in Section \ref{sec: siamese}, $L_R$ is the reconstruction of the \textit{anchor, positive}, and \textit{negative} terms, and $L_C$ corresponds to the configuration approximation for the 3 terms as well:

\[ L_R(x_a^i, x_a^k, x_j^i) = ||\tilde{x}_a^i -  x_a^i||^2 + ||\tilde{x}_a^k -  x_a^k||^2 + ||\tilde{x}_j^i -  x_j^i||^2 \]

\[ L_C(v_a^i, v_a^k, v_j^i) = ||\tilde{v}_a^i -  v_a^i||^2 + ||\tilde{v}_a^k -  v_a^k||^2 + ||\tilde{v}_j^i -  v_j^i||^2 \]

\textbf{Hybrid2. } In contrast to the triplet loss that samples one positive and one negative point
for each anchor point in a batch of data, 
 the \textbf{Soft Nearest Neighbor} (SNN) loss ~\cite{Frosst2019AnalyzingAI,pmlr-v2-salakhutdinov07a} uses all the points in the batch to measure the separation between classes. We apply this loss to an encoder layer of the autoencoder so that the joint loss that we minimize has
 both the reconstruction term and the SNN term.

\[ \mathcal{J} = \displaystyle\frac{1}{N} \sum_{i, j}\left( ||\tilde{x}_j^i -  x_j^i||^2 - \lambda~\text{log}\left(\frac{\displaystyle\sum_{k\neq i}e^{-\frac{||z_j^i - z_j^k||^2}{T}}}{\displaystyle\sum_{\substack{k,l \\ (k,l) \neq (i,j)}}e^{-\frac{||z_j^i - z_l^k||^2}{T}}}\right) \right)\]
where $\lambda$ is a regularization coefficient and $T$ is a temperature hyperparameter.

The soft nearest neighbor term for one training point (represented by $i$ as index for configuration and $j$ as index for job) is given by (assuming $T=1$ for now):
\[  - ~\text{log} \frac{\displaystyle\sum_{k\neq i}exp(-||z_j^i - z_j^k||^2)}{\displaystyle\sum_{\substack{k,l \\ (k,l) \neq (i,j)}}exp(-||z_j^i - z_l^k||^2)}
= -~\text{log}\frac{numerator}{denominator}\]

The numerator is a sum of negative exponentials of distances between the encoding $z_j^i$ of the current job $j$ with the current configuration $i$ and all other encodings $z_j^k$ for the same job $j$ within the same batch but obtained under a configuration $k$ different than the initial configuration $i$ (hence $k\neq i$) (so it's a sum of distances between all "positive pairs"). The denominator is a sum of negative of exponentials of distances between the encoding $z_j^i$ and all other encodings $z_l^k$ coming from different jobs $l$ (hence $l\neq j$) and under different configuration from the current $i$ (hence $k\neq i$).

We are minimizing the soft nearest neighbor term, which is equivalent to minimizing the denominator 
and maximizing the numerator
because $log$ is a monotonically increasing function.
The numerator is a sum over positive terms. We can maximize it by maximizing each of its term. Maximizing $exp(-distance)$ is equivalent to minimizing the distance within the exponential term. So we are trying to minimize the distance between encodings coming from the same workload ($z_j^i$ and $z_j^k$).
On the other hand, we are also minimizing the denominator which is as well a sum of positive terms. Minimizing the denominator consists of minimizing each term.  Each term is minimized if the distance inside the negative exponential is maximized. So this corresponds to maximizing the distance between the current encoding $z_j^i$ and other encodings $z_l^k$ coming from different workloads under a different configuration.

As for the temperature parameter, it controls the "radius of neighbourhood" regarding points within the batch to take into account within both the numerator and the denominator.
If T is very high (close to infinity), then the values of the distances will not be taken into account. Instead, the numerator will be the number of points within the same batch that belong to the same workload, and the denominator will be the number of points within the same batch that belong to a different workload. This means that under very high values of T, minimizing this loss function is not useful for learning representations that are more tightened if they belong to the same workload but far apart if they belong to different workloads.
If T is very low, then the loss becomes extremely sensitive to distances between the points and a small change in the distance can make a big difference in the value of the SNN function.
So T can be seen as a \textit{smoothing} parameter.

\mysection{Experimental Results}
\label{sec:expt}
\label{headings}

In this section, we evaluate all of our  modeling methods using benchmark data. 

\mysubsection{Benchmarks and Trace Collection}
We developed two benchmarks of Spark workloads based on (1)~an extension of the streaming workloads from prior work~\cite{LiDS15}
and (2)~TPCx-BB \cite{tpcxbb}.
We collected a {\em trace} for each workload under a particular configuration, covering two types of  metrics: ($i$)~Spark related metrics, collected within the Spark listener; and ($ii$)~OS related metrics,  collected using the unix command \textit{nmon}.

Both benchmarks cover a wide range of analytics, ranging from SQL queries to ETL tasks (using SQL  and UDFs)  to ML tasks.
The streaming  benchmark tunes 10 knobs (dimension of $v_j^i$ is 10) and comprises 70 workloads, including 53 training workloads and 17 test workloads, with 128 traces each. 
The TPCx-BB  benchmark tunes 12 knobs (dimension of $v_j^i$ is 12) and includes 30 templates, from which we generated 1160 workloads via parameterization. 
Among them, 928 are used as training workloads,  including ($i$) 58 intensively sampled workloads with around 315 traces each,  which represent the special workloads  made available by the application for offline sampling, ($ii$) 870 sparsely sampled workloads  with around 30 traces each, representing online user workloads with fewer configurations observed. Finally, 232 are reserved as test workloads, with 30 traces each.

\textbf{Preprocessing.}
In each trace, we take the average of the metrics across the execution period of the
Spark workloads and then minmax-scale both the runtime metrics $x_j^i$ as well
as the configuration knobs $v_j^i$. We drop constant metrics and end up with 561 metrics for each streaming workloads trace and 286
metrics for each batch workloads trace.
The preprocessed traces alongside our code are available at:
\url{https://github.com/udao-modeling/code}

\mysubsection{Evaluation Methodology}

We provide the main comparative results between different modeling techniques in Table~\ref{table2} . We start with results from a baseline called "\textit{all metrics}" and that bypasses representation learning and uses the whole vector of trace $x_j^i$ as the encoding for the workload ($z_j^i = x_j^i$ in this case). Then, we introduce two other baseline methods from the early literature of representation learning: \textit{PCA} and \textit{KPCA} and use them as an encoding extraction tool instead of neural based auto-encoders.  Then, we list the results obtained with the previously introduced neural network modeling techniques: (1)  \textit{Embedding} architecture introduced in \S \ref{section: embedding-approach}, (2) \textit{Custom autoencoder}, (3) \textit{Custom contractive autoencoder} and (4) \textit{Variational autoencoder}  from \S \ref{vae-paragraph}, and (5) the \textit{siamese neural network} from \S \ref{sec: siamese}. We also list results from the 2 \textit{hybrid} methods introduced in \S \ref{section-hybrid-architectures}. Finally, we compare to a state of the art tuning tool, Ottertune~\cite{ottertune,ottertune-demo}.

\textbf{Encoding Extraction Scheme.}
We consider two schemes for extracting encodings from configurations:
($a$) \textit{shared scheme}:  $z_j$ is extracted from traces coming from a shared pool of
configurations (averaging $\{z_j^i\}_{i}$ with $i$ selected from the shared
pool).
($b$) \textit{arbitrary scheme}: $z_j$ is extracted from traces coming from an arbitrary pool
of configurations (averaging $\{z_j^i\}_{i}$ with $i$ selected from the arbitrary
pool).
We also distinguish between extracting the encoding for test workloads with either
1 or 5 observations, under each of the above ($a$) and ($b$) schemes, 
as shown in the header of Table \ref{table2}.

It is worth noting that the \textit{arbitrary scheme} is more practical than the \textit{shared scheme} since a cloud optimizer must expect receiving an \textit{arbitrary} configuration for a newly submitted job. The modeling problem becomes even harder when only 1 trace is observed for the test workload. Nevertheless, we explicitly make the comparison between the two schemes in Table \ref{table2} to better understand which modeling technique works best under different job admission settings, and we color the most practical case \textit{(arbitrary, 1 ob)} in Table \ref{table2} %. \YD{Use a darker color}

\textbf{Evaluation Metric.} We use the Mean Absolute Percentage Error  (MAPE) metric
for reporting results for the different modeling methods.

\textbf{Hyper-parameter tuning.}
For the encoder/decoder based architectures as well as the neural networks
we tune topology, optimization and other hyperparameters (such as coefficients within loss functions)
by using a 5 fold cross validation scheme that simulates the same training settings as in practical cases
(observing few configurations for workloads in the left out fold).

\textbf{Hardware and Implementation Details.}
Our workloads are deployed on several Spark clusters, each spanning 1 node for the driver and 2 for the executors.
Each node has 2 processors (\textit{Intel(R) Xeon(R) Gold 6130 CPU @ 2.10GHz})
totalling 32 cores and 754 GB of RAM for each node.
The modeling approaches have been implemented mainly in Tensorflow\cite{tensorflow2015-whitepaper},
Keras\cite{keras} and scikit-learn\cite{sklearn}.

\begin{figure*}[t]
	\vspace{-5mm}
	%\begin{figure*}
	\includegraphics[scale=.3]{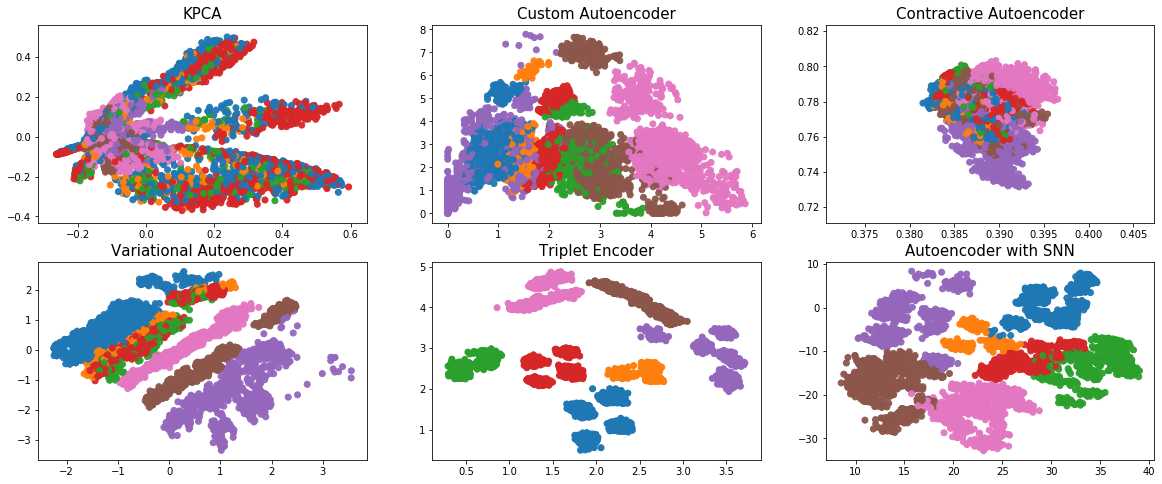}
	
	\caption{\small 2D encodings obtained with different encoding/decoding techniques using the streaming trace dataset. \textit{Different colors represent different templates of workloads.}}
	\label{fig:visualizations}
	%\end{figure*}
	\vspace{-2mm}
\end{figure*}

\begin{table*}[t]
%	\vspace{-2mm}
	\centering
\begin{adjustbox}{width=1\textwidth}
	\begin{tabular}{c|ccca|ccca}
		\toprule 
		 & \multicolumn{4}{c}{\textbf{Streaming Trace}} &  \multicolumn{4}{c}{\textbf{TPCx-BB Trace}} \\ 
		& \multicolumn{2}{c}{\textbf{Shared Pool \,\,\,}} &  \multicolumn{2}{c|}{\textbf{Arbitrary Pool}} 
		& \multicolumn{2}{c}{\textbf{Shared Pool \,\,\,}} &  \multicolumn{2}{c}{\textbf{Arbitrary Pool}} \\ 
		 & \textbf{5 obs} &\textbf{ 1 ob} & \textbf{5 obs} & \textbf{\textcolor{black}{1 ob}}
		 & \textbf{5 obs} &\textbf{ 1 ob} & \textbf{5 obs} & \textbf{\textcolor{black}{1 ob}} \\ 
		\midrule
		%	   Baselines for representation learning
All metrics (scaled) & 11.9 & 10.9 & 34.8 & 34.9
& \textbf{7.2} & \textbf{7.6} & 8.6 & 29.6 \\ % results from 2020-06-05

PCA & 11.4 & 11.3 & 24.4 & 60.7
& 11.9 & 16.4 & 70.1 & 50.8 \\ % results from 2020-06-05
KPCA &\textbf{8.5}& \textbf{9.9} & 17.9 & 21.3
& 35.2 & 42.8 & 59.0 & 58.8 \\ % results from 2020-06-02
\midrule

Embedding & 32.8& - & 22.5& \textcolor{black}{-} % results from 2020-05-24; commit:
& 14.7 & - & 12.4 & \textcolor{black}{-} \\ % results from 2020-06-02
\midrule

%		\midrule
		Custom AE & 16.0& 13.0 & 20.2 & 21.4 % results from 2020-01-28  report 
				            & 16.9 & 22.2 & 19.2 & 49.9 \\  % results from 2020-06-02
%		 so called contractiveplus previously
Custom contractive AE& 10.6 & 12.2 &  13.0 & 19.7 % results from 2020-05-12 report
& 9.7& 14.3 & 28.9 & 53.0 \\  % results from 2020-06-02 & 2020-06-05
		VAE & \textbf{8.5} & 11.2 & 17.7 & 18.7 % results from 2020-05-12 report 
				            & 11.4 & 14.6 & 28.4 & 37.5 \\  % results from 2020-06-02
%		\midrule
		Siamese Network (triplet) & 10.6 & 12.6 & \textbf{9.6} & \textbf{11.6}
				             & 7.7 & 7.9 & 6.5 & \textbf{9.5
				             } \\  % results from 2020-06-02
		\midrule
%		This is what I called before SNN
%		\midrule

        Hybrid1 & 11.4& 12.0 & 27.0 & 11.9 %results from 2020-05-25; commit=43a60efa96
        & \textbf{7.6 }& 8.2 & \textbf{6.2} & 9.7 \\  % results from 2020-06-02
        Hybrid1($\lambda=0$) & 10.3& 11.5 & 10.5 & 12.6 % results from 2020-04-23 report
        &  \textbf{7.6} &\textbf{7.6} & 6.3 & 9.6 \\  % results from 2020-06-02
        Hybrid2 & 9.9 & 12.4 & 11.2 & 12.8 % results from 2020-05-24; commit: d2390e254
        & 7.9  & 8.3 & 6.8 & 10.7 \\  % results from 2020-06-02
		\midrule
Ottertune (default) & 83.7 & 84.0 & 67.6  & 95.5 % results from this commit: d513e424 (Mar 22nd)
& 52.1 & 44.6 & 42.2 & 61.2 \\  % results from 2020-06-02
Ottertune (tuned) & 50.8 & 63.8 & 36.8 & 67.8 % results from this commit: d513e424 (Mar 22nd)
& 41.0 & 33.5 & 35.2 & 38.2 \\  % results from 2020-06-02

		\bottomrule
	\end{tabular}
\end{adjustbox}
	\caption{Runtime latency MAPE computed over test sets and averaged over 10 runs}
	\label{table2}
%\vspace{-0.1in}
\end{table*}

\mysubsection{Comparative Results of Modeling Techniques} 
\label{sec53}

We make the following observations from Table~\ref{table2} and profiling results in Fig.~\ref{fig:visualizations}:

1. \textbf{Baseline Methods.} If we bypass representation learning techniques and directly train a global regressor model on the runtime metrics $x_j^i$ (but taking their job centroid $x_j$) alongside the input configuration(s) $v_j^i$, then we can get low errors on the  latency estimation if we guarantee having seen a job configuration from the \textit{shared pool}.
Similarly, \textit{PCA} and \textit{KPCA}, two basic representation learning techniques, also work well under the same \textit{shared scheme.}
These baseline methods, however, fail  to work when a job is admitted by the system under an \textit{arbitrary} configuration.
A closer look at the encodings obtained with \textit{KPCA} applied on raw metrics $x_j^i$ in Figure \ref{fig:visualizations} shows how encodings from different job templates are scattered in the 2D space and thus clearly violate the invariance property.

2. \textbf{Autoencoders.}
The \textit{custom autoencoder} fails to provide better performances than baseline
methods under the different schemes. Its design, which mainly focuses on
reconstructing the variant part by adding a supervision term to the
reconstruction loss function, fails to offer the invariance property in the other designated
part of the bottleneck layer. This insight is verified in Figure
\ref{fig:visualizations}: while encodings learned from the custom autoencoder have better
clustering properties, according to different jobs, than those learned from a
basic autoencoder or KPCA, they are still scattered and not tight enough
along each job's centroid.

Further adding a \textit{contractive} term on top of our custom autoencoder provides
consistently better results across all encoding schemes for the streaming trace,
but only under shared scheme for the TPCx-BB trace. The \textit{contraction} is induced by adding the Frobenius
norm of the Jacobian matrix in Eq.~\ref{eq:contractive-ae}. This additional unsupervised term hence
doesn't condition the invariance of encodings according to each specific workload, but rather affects
all workloads encodings by contracting them at once as seen in Figure \ref{fig:visualizations}.

On the other hand, the \textit{variational autoencoder} further improves the
errors on the streaming trace, but doesn't bring improvements on the
TPCx-bb trace, especially when it comes to the \textit{arbitrary scheme}. 
By examining the encodings obtained from this approach in Figure \ref{fig:visualizations},
we see similar clustering properties as the one induced by our custom disentanglement.

3. \textbf{Siamese neural networks} focus on a relaxation of the invariance property
and achieve drastic improvements on the errors obtained in the most constrained (challenging) setting 
of observing 1 \textit{arbitrary} configuration for an admitted job,  under both streaming and TPCx-BB
datasets.
The success of this architecture is attributed to its capacity to tightening
encodings from traces of the same workload and separating encodings of different
workloads, and thus focusing on learning a more invariant encoding for each
workload.

4. \textbf{Hybrid methods.}
Augmenting the triplet loss function with a reconstruction term and a custom
disentanglement didn't bring improvements beyond those achieved with the siamese
neural network alone.  Indeed, while tuning the hyperparameters of the loss function in Hybrid1,
we found that $\gamma$ was assigned a small value for the best hyperparameters
 chosen, which indicates that the loss function puts less emphasize
on the reconstruction term.
Further, by closely examining the results obtained with Hybrid1 and
Hybrid1 ($\lambda=0$), we can conclude that the invariance property subsumed
independence in our problem settings across the two datasets. The supervised
triplet loss function gave indeed consistent results on the test sets no matter
how many (1 or 5) and from which pool (\textit{arbitrary} or \textit{shared})
configurations were sampled.
The second hybrid loss function, which combines a soft nearest neighbor term  
(a more recent metric learning method) and a reconstruction term, provides error
on the same scale as the first hybrid loss function with $\lambda$ set to 0.

5.\textbf{Ottertune}\cite{ottertune,ottertune-demo}, a state-of-the-art tuning tool for RDBMS, does not leverage traces from
different workloads or use representation learning techniques to train a single model. In contrast to our approaches, it trains one model
per workload and then maps each test workload to one of the past training workloads in order to model its performances.
This leads to higher errors across the different training settings under both datasets.

6. \textbf{Embedding.} The \textit{embedding} approach we introduced earlier in section \ref{section: embedding-approach} doesn't fully use the raw metrics $x_j^i$
to extract an encoding upon the admission of a new job. Instead, it learns an embedding by backpropagating the least squares loss
that focuses solely on the actual runtime latency $y_j^i$ to learn a unique encoding $z_j$. Although this approach fully satisfies the invariance property, it remains inferior to other neural based approaches grounded in representation learning. This is because it leverages less information while learning the workload encoding. Despite that fact, it still outperforms Ottertune.
Since the embedding approach requires incremental training before being able to predict, it requires a number of observed data points at least equal to the degrees of freedom (the number of components) of the embedding vector. Therefore, we don't apply this approach when having only 1 observation.

\mysubsection{End-to-End Experiments} 
\begin{figure*}
	\vspace{-5mm}
	\includegraphics[scale=.39]{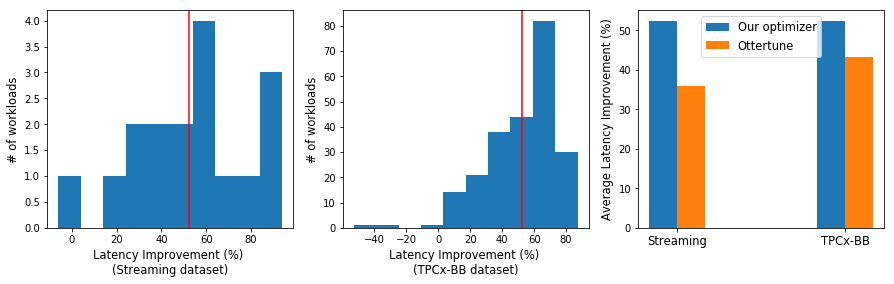}
	
	\caption{\small End to end performances and comparison to Ottertune.}
	
	\label{fig:end-to-end}
	\vspace{-2mm}
\end{figure*}
\vspace{-5mm}
We use our best modeling technique (siamese neural network) in order to drive an end-to-end tuning experiment while observing a single arbitrary configuration for each test job. This initial arbitrarily set configuration is not necessarily the same across different test workloads.
 We then run an %exhaustive search
  optimizer that enumerates combinations of different knob choices, each with a predicted latency by the siamese neural network, and recommends a configuration to minimize the latency of each test job. We record the runtime latency for the recommended configuration\footnote{\footnotesize The optimizer's recommendation is sometimes too optimistic due to extrapolation in a sparse search space. If the job fails to be launched with the recommended configuration, the optimizer recommends another one.}, and then compute the average of latency improvement over the initial configuration,  ($1 - \frac{\text{new latency}}{\text{initial latency}}$), across the different workloads.
  
Figure \ref{fig:end-to-end} gives us direct insights on the distribution of speedup recorded for the runtime latency of workloads from both benchmarks. The left and center plots show histograms for average speedups on test workloads from both datasets. 
  The rightmost plot shows average latency improvements obtained with our method and the ones obtained by Ottertune \cite{ottertune}.
  On average, we achieve a latency improvement of $52.4\%$ on streaming workloads and $52.44\%$ on TPCx-BB workloads, compared to $35.96\%$ and $43.19\%$ for Ottertune, respectively.
  
  After closely examining the configurations recommended by our method and Ottertune, we noticed that both methods aggresively increase the amount of resources allocated in most of the test workloads. Increasing the amount of resources allocated for a workload (such as the total number of cores and the memory per executor) yields in general better runtime latencies regardless of the choice of the remaining knobs. This explains why the gap is not very big between both methods when it comes to end-to-end performances. However, in some of our test workloads, where initial configurations are already assigned the biggest resource capacity, and where both optimization methods keep the resource allocation knobs intact but change other knobs, our method tends to recommend better configurations than Ottertune.

\label{others}

\mysection{Conclusions and Future Work}
In this paper, we presented our solution to performance modeling for cloud data analytics, including 
($i$)  a system design that suits the constraints in real world applications, ($ii$) a notion of learning 
workload embeddings with desired properties  for different jobs,  thereby  enabling  performance 
prediction when used together with job configurations; 
($iii$)  an in-depth study of different modeling choices that meet our requirements. 
Results of  extensive experiments show the strengths and limitations of different modeling methods, 
reveal the best performing technique to be the one that can best approximate the invariance property of 
workload embeddings, and demonstrate our superior performance over a state-of-the-art modeling 
technique for cloud analytics.
In future work, we plan to extend our analytics model server with transfer learning capabilities 
to  efficiently learn performance models on different hardware types, and more advanced workload 
embedding techniques that can leverage logical descriptions, such as SQL query plans, that are 
available to a subset of workloads.

%\newpage

%\end{flushleft}
%\renewcommand{\baselinestretch}{0.8}
 \bibliographystyle{splncs04}
\small
 \bibliography{mybib,bigdata}
% \input{8-impact}
% \input{9-appendix}
%\renewcommand{\baselinestretch}{1}
%Author, F.: Article title. Journal \textbf{2}(5), 99--110 (2016)
%Author, F., Author, S.: Title of a proceedings paper. In: Editor,
%F., Editor, S. (eds.) CONFERENCE 2016, LNCS, vol. 9999, pp. 1--13.
%Springer, Heidelberg (2016). \doi{10.10007/1234567890}
%Author, F., Author, S., Author, T.: Book title. 2nd edn. Publisher,
%Author, A.-B.: Contribution title. In: 9th International Proceedings
%on Proceedings, pp. 1--2. Publisher, Location (2010)
%LNCS Homepage, \url{http://www.springer.com/lncs}. Last accessed 4
%\end{thebibliography}

\end{document}